\documentclass[journal=nalefd,manuscript=article]{achemso}

\usepackage{comment}
\usepackage{color}
\usepackage{graphicx}% Include figure files
\usepackage{subfigure}
\usepackage{dcolumn}% Align table columns on decimal point
\usepackage{bm}% bold math
\usepackage{amsmath}
 
\usepackage{braket}
\usepackage{diagbox, eqparbox, hhline}
\usepackage[detect-none]{siunitx}
\usepackage{ulem}
\usepackage{mathtools}

\DeclareMathAlphabet{\mathpzc}{OT1}{pzc}{m}{it}

\newcommand\myeq{\stackrel{\mathclap{\footnotesize\mbox{def}}}{=}}
\usepackage{chemformula} % Formula subscripts using \ch{}
\usepackage[T1]{fontenc} % Use modern font encodings

\author{Denis \v{S}abani}
\affiliation{%
Departement Fysica, Universiteit Antwerpen, Groenenborgerlaan 171, B-2020 Antwerpen, Belgium
}
\alsoaffiliation{%
NANOlab Center of Excellence, University of Antwerp, Belgium
}

\author{Cihan Bacaksiz}
\alsoaffiliation{%
Departement Fysica, Universiteit Antwerpen, Groenenborgerlaan 171, B-2020 Antwerpen, Belgium
}
\affiliation{%
NANOlab Center of Excellence, University of Antwerp, Belgium
}

\author{Milorad V. Milo\v{s}evi\'c}%
\email{milorad.milosevic@uantwerpen.be}
\affiliation{%
Departement Fysica, Universiteit Antwerpen, Groenenborgerlaan 171, B-2020 Antwerpen, Belgium
}%
\alsoaffiliation{%
NANOlab Center of Excellence, University of Antwerp, Belgium
}

\title[An \textsf{achemso} demo]
  {Releasing latent chirality in magnetic two-dimensional materials}

\abbreviations{IR,NMR,UV}

\begin{document}


\begin{abstract}
Dzyaloshinskii-Moriya interaction (DMI) is at heart of chiral magnetism and causes emergence of rich non-collinear and unique topological spin textures in magnetic materials, including cycloids, helices, skyrmions and other. Here we show that strong intrinsic DMI lives in recently discovered van der Waals magnetic two-dimensional (2D) materials, due to the sizeable spin-orbit coupling on the non-magnetic ions. In a perfect crystal, this intrinsic DMI remains hidden, but is released with any break of point-inversion symmetry between magnetic ions, unavoidable at the sample edges, at ever-present structural defects, with any buckling of the material, or with non-uniform strain on an uneven substrate. We demonstrate such release of latent chirality on an archetypal magnetic monolayer CrI$_{3}$, and discuss the plethora of realizable DMI patterns, their control by nanoengineering and tuning by external electric field, thereby opening novel routes in 2D magnetoelectronics.
\end{abstract}

\bigskip

Recent realizations of the 2D ferromagnetic materials, monolayer CrI$_3$\cite{huang2017layer} and bilayer Cr$_2$Ge$_2$Te$_6$ \cite{gong2017discovery}, opened the gate for many magnetic 2D materials synthesized since. In spite of the Mermin-Wagner theorem, 2D magnetism in these materials is possible, even at finite temperatures, due to the magnetic anisotropy stemming from the strong spin-orbit coupling (SOC) on the non-magnetic ions. Although the phase transition temperatures are in general relatively low, 2D magnetism can be sustained in some materials even up to room-temperature, such as reported for monolayer VSe$_2$ \cite{bonilla2018strong} and MnSe$_2$ \cite{o2018room}. All the experimental evidence to date suggests that in pristine 2D magnetic materials the spins align parallel to the easy-axis direction and hence show no apparent magnetic chirality.

The crucial ingredient for chiral magnetism to arise is the Dzyaloshinskii-Moriya interaction (DMI) \cite{dmi_moriya,dmi_dzyal}. Such interaction appears in materials with strong spin-orbit coupling which lack point inversion symmetry between the magnetic ions, and favors orthogonal ordering of adjacent spins, thereby competing with the symmetric magnetic exchange that aligns the spins. This competition results in rich noncollinear spin textures, among which skyrmion lattices \cite{muhlbauer2009skyrmion,yu2010real,fert2017magnetic} and spin spirals \cite{bode2007chiral}. Previously, DMI was found to emerge due to symmetry breaking in non-centrosymmetric bulk materials \cite{dmitrienko2014measuring,beutier2017band,miyawaki2017dzyaloshinskii,du2015edge} and in ultrathin elemental ferromagnets (Fe, Co, Ni) interfaced with a heavy-metal layer \cite{cho2015thickness,ma2018interfacial,romming2015field,sampaio2013nucleation,PhysRevLett.118.219901} or graphene \cite{yang2018significant}. However, DMI did not receive much attention in recently discovered two-dimensional (2D) magnetic materials\cite{huang2017layer,gong2017discovery,gibertini2019magnetic}. Besides the fact that such materials are still very new, they do preserve inversion symmetry, and their magnetic atoms (in latter case chromium) can hardly be sensitive to interfacing by a heavy-metal layer due to nonmagnetic layers in between (in latter case iodine) - hence it is not intuitive that any sizeable DMI can be induced in them using the established routines.

However, one quickly recalls that it is the strong spin-orbit coupling in the nonmagnetic ligands in latter 2D materials that supports appearance of long-range magnetization in spite of the Mermin-Wagner theorem \cite{lado2017origin,XuKitaev,huang2017layer,gong2017discovery}. If so, then one expects that spin-orbit coupling on the non-magnetic ligands can also prompt the DMI to arise. It is definitely plausible that each non-magnetic ion contributes some DMI to a pair of magnetic ions, but the net DMI stemming from all nearest-neighbor non-magnetic ions will cancel out in a perfect lattice. However, once the point-inversion symmetry between the magnetic atoms in a 2D magnet is broken, this \textit{latent} DMI should reveal itself. 

Intuitively, one does not expect such DMI to be large. However, very recently it was predicted that symmetry-breaking in the Janus (two-faced) forms of CrX$_3$ (X= I, Br, and Cl) and MnX$_2$ (X=S and Se) can produce strong DMI, sufficient to obtain skyrmion-like spin textures, even at finite temperatures \cite{PhysRevB.101.060404,Liang2019VeryLD,Yuan2019ZerofieldSS}. Still, Janus 2D materials are far from trivial to realize with the current experimental state-of-the-art. Instead, every synthesized or exfoliated 2D magnetic material, on a substrate, will exhibit enough structural deformations to warrant local release of latent DMI. Understanding and control thereof is the primary objective of this Letter. 

Specifically, we demonstrate the release of latent DMI in monolayer CrI$_{3}$, hidden by the point-inversion symmetry between the nearest-neighbor Cr ions. We tackle examples of symmetry breaking that are ever present in experimental samples, such as lateral edges, inhomogeneous strain, buckling, vacancies, or substitutional atoms, and external manipulations such as the applied electric field.\cite{EfieldDMI,EfieldDMI1,EfieldDMI2,EfieldDMI3} Finally, we go on to quantify the link between the structural and electronic manipulation of the DMI release, and provide guidelines on how chirality can be induced and tailored in 2D magnetic materials. Ability of such controlled release of latent DMI is of clear fundamental value, but also bears relevance to a spectrum of potential applications in magnonics and spintronics.\cite{magnonics_ref,spintronics_ref}

To describe the magnetic interactions, we consider Heisenberg spin Hamiltonian $H=\frac{1}{2}\sum_{i , j} \mathbf{S}_i \mathbf{J}_{ij} \mathbf{S}_j + \sum_{i} \mathbf{S}_i \mathbf{A}_{ii} \mathbf{S}_j$, where $\mathbf{S}_{i} = (S_{i}^{x},S_{i}^{y},S_{i}^{z})$ is the spin vector of the $i$th site. $\mathbf{J}_{ij}$ and $\mathbf{A}_{ii}$ are $3 \times 3$ matrices describing the total magnetic exchange interaction between the different sites and the single-ion anisotropy (SIA), respectively. Total exchange consists of the symmetric and the antisymmetric (DMI) contribution. DMI energy contribution can be written as $E_{DM} = \sum_{(i,j)}\mathbf{D}_{ij}\cdot (\mathbf{S}_{i}\times\mathbf{S}_{j})$, where $\mathbf{D}_{ij} = (D_{ij}^{x}, D_{ij}^{y}, D_{ij}^{z})$ is the DMI vector between magnetic ions on $i$th and $j$th site. In this Letter, we will focus our attention on the antisymmetric (DMI) exchange, as spin-aligning symmetric exchange and single-ion anisotropy were given enough attention in earlier works.\cite{lado2017origin,XuKitaev} The DMI vector between the nearest-neighbor magnetic ions is extracted from the total exchange matrix between them using the $D^{\gamma}_{ij} = \frac{1}{2}\cdot (J^{\alpha\beta}_{ij}-J^{\beta\alpha}_{ij})$, where $(\alpha,\beta,\gamma)$ are cyclic Cartesian coordinates $(x,y,z)$, $(y,z,x)$ and $(z,x,y)$. In order to obtain total exchange matrix elements, and consequently calculate the profile of DMI vectors between the magnetic ions in different structures based on monolayer CrI$_{3}$, we apply the Four-State Methodology (4SM)\cite{4sm2011,4smDT2013,4smarXiv} based on mapping \textit{ab initio} energies onto given Heisenberg spin Hamiltonian. 

\subsubsection{Magnetic chirality at iodine vacancies} \label{vac}

\begin{figure*}[ht]
\includegraphics[width=0.7\linewidth]{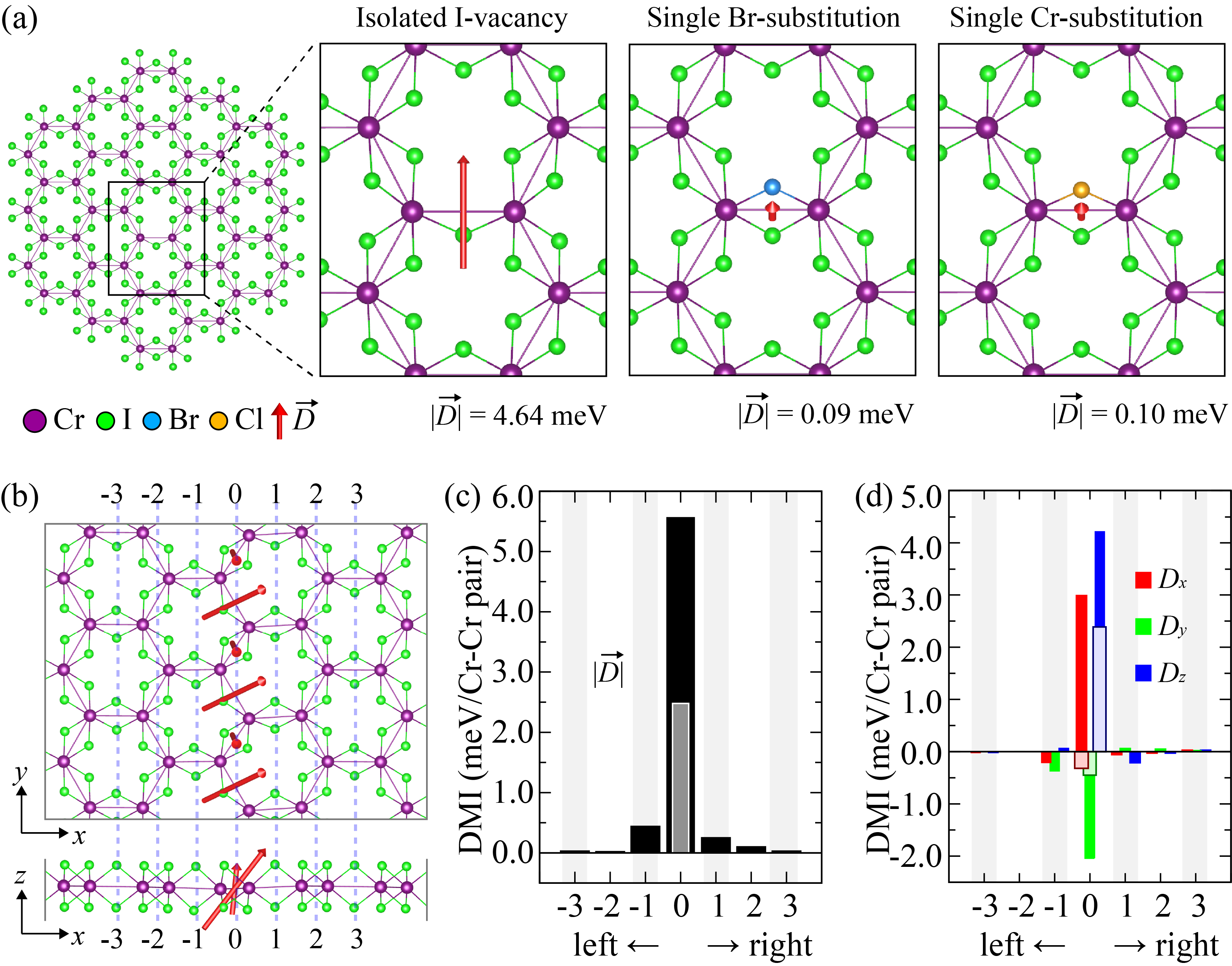}
\caption{\label{ff1}
\textbf{(a)} The released DMI at an isolated I vacancy (left), a substitution of Br atom (middle), and a substitution of Cl atom (right). \textbf{(b)} The profile of awaken DMI (red arrows) in case of a line of iodine vacancies in the monolayer CrI$_{3}$, top (above) and side view (below). Only the strongest DMI vectors are shown. \textbf{(c)} Distribution of the DMI intensity in the vicinity of the defect line. \textbf{(d)} Distribution of the Cartesian components of DMI, illuminating the change in direction of the DMI vector around the defect.}
\end{figure*}

In either exfoliated or directly synthesized 2D materials, the structural imperfections such as atom vacancies are unavoidable. In case of chromium halide monolayers, the vacancies in the outer halide layers are the most likely ones to appear, which obviously breaks the inversion symmetry for the chromium atoms in the vicinity of the missing atom(s). Hence such a realistic situation is clearly favorable for the appearance of DMI. We therefore start our consideration from an isolated, single iodine vacancy in otherwise pristine CrI$_3$ monolayer, as shown in Fig.~\ref{ff1}a. We found that the arising DMI is largest at the Cr-Cr pair that misses the iodine ligand, with the DMI vector orthogonal to the Cr-Cr bond\cite{dmi_moriya}. The intensity of such DMI, created at an isolated single iodine vacancy, is surprisingly large - $4.64$ meV/Cr-Cr pair - which is comparable to artificially created DMI at Co/Pt interfaces.\cite{PhysRevLett.118.219901} 

During the formation of the crystal the defects are often not isolated from each other, but rather form a pattern, such as e.g. dislocation or grain boundary. Moreover, the recent discoveries report that ligand vacancies can be deliberately created in transition-metal complexes, such as monolayer WS$_{2}$.\cite{sulf_vacancies2019} Therefore, for comparison to the isolated vacancy case, we have also investigated a line of iodine-vacancies in monolayer CrI$_{3}$. Due to computational limitations, the considered line-defect was ideally parallel to the zig-zag chain of Cr atoms, as shown in Fig.~\ref{ff1}b. Although we graphically present only the DMI arising on that chain of Cr atoms (being dominant), the line-defect will awake weak yet non-negligible DMI in the broader neighborhood as well. It exists in a narrow region, ranging less than 1 nm from the vacancies, and exhibits direction that deviates from being orthogonal to corresponding Cr-Cr bond by the small angle. The spatial distribution of the DMI intensity, and its vectorial components, are shown in Figs. \ref{ff1}b and \ref{ff1}c, respectively, for different Cr-Cr pairs sequentially away from the line-defect. The intensity of DMI for the pair missing one ligand is 5.57 meV/Cr-Cr pair; the pair with both ligands present, but situated on defect line experiences DMI of 2.48 meV/Cr-Cr pair. The DMI on the first-nearest pairs on the left and the right side of the defect line falls to 0.44  and 0.25 meV/Cr-Cr pair, respectively, thus decays by an order of magnitude compared to the DMI on the defect line. The trend continues at the second-nearest pair on the right side, where DMI amounts to 0.1 meV/Cr-Cr pair, while at all the other pairs, further from the defect line, DMI can be considered dormant (as in the pristine CrI$_{3}$ monolayer). DMI liberated at iodine vacancies in monolayer CrI$_{3}$ has intensity comparable to the highest values known in artificial magnetic heterostructures. We note that the isotropic exchange interaction increases twice at the defect site.

\subsubsection{Magnetic chirality caused by ligand substitution} \label{subs}

Having understood the case of iodine vacancy, we next look at the DMI awaken by substitution of a foreign ligand to the iodine site. As two plausible cases, we chose substitution of I by Br or Cl, which are also halides, knowing that Cr-Br and Cr-Cl compounds also exhibit magnetism\cite{ghazaryan2018magnon,klein2019enhancement}. As a first result, we note that substitutional (either Br or Cl) atom resides much closer to the Cr-Cr pair than iodine atom did, which already breaks the inversion symmetry and suggests that DMI should appear. Indeed we found DMI on the Cr-Cr pair of 0.09 and 0.10 meV for Br and Cr substitution, respectively. As shown in Fig.~\ref{ff1}a DMI vector is orthogonal to the Cr-Cr bond pointing the substituted atom with a very slightly rotation from the Cr-I-Cr-X plane (X=Br or Cl) through the in-plane direction. The DMI values here are consistent with recently reported values for Janus monolayers of Cr(I,Br)$_{3}$ and Cr(I,Cl)$_{3}$, being 0.27 meV/Cr-Cr pair and 0.19 meV/Cr-Cr pair, respectively.\cite{PhysRevB.101.060404} However, they are ~50 times smaller as compared to DMI at an isolated iodine vacancy. The substitution of a foreign ligand completes the octahedral coordination of the Cr pair and thereby recovers the exchange pathways as in pristine CrI$_3$, minimizing the DMI. Still, the differences in the structure as well as in the spin-orbit coupling of the participating atoms and the electronic structure of the resultant crystal strongly affect the magnetic interaction and cause DMI to appear.

\subsubsection{Chirality at sample edges} \label{edge}

\begin{figure*}[ht]
\includegraphics[width=0.60\linewidth]{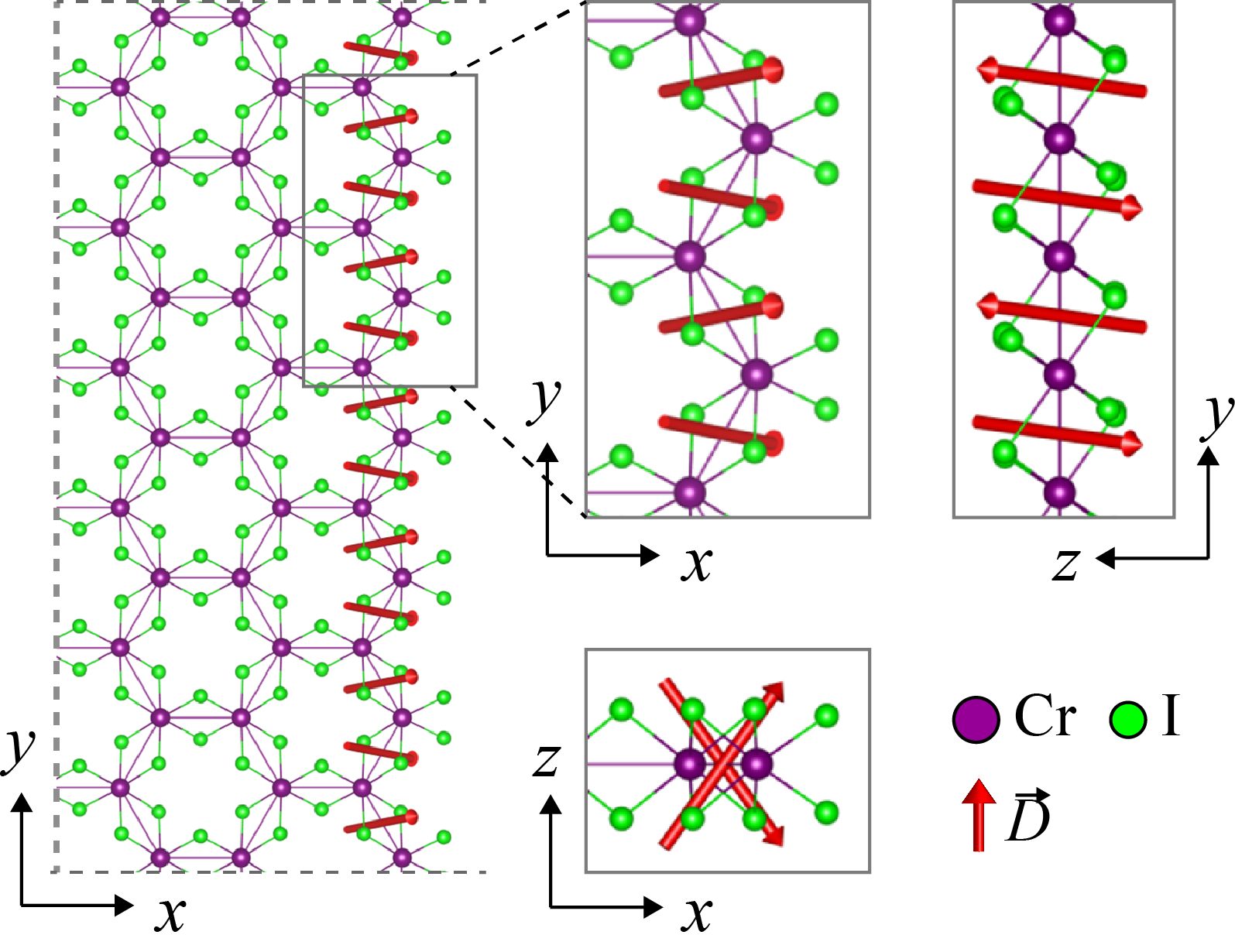}
\caption{\label{ff2}
The profile of awaken DMI (red arrows) near the edge on the monolayer CrI$_{3}$, top (above) and side view (below).}
\end{figure*}

The symmetry-breaking at the boundaries of any magnetic film is expected to cause the so-called boundary-induced chirality, completely
independent of any internal DMI field.\cite{edgeDMI1,edgeDMI2} In the past, such contribution to DMI was derived using particular tensorial constructions with respect to the point group of the system. In the case of 2D materials however, the DMI arising at the edges can be evaluated directly from first principles. With that goal, we examined the zig-zag edge of a monolayer CrI$_3$ nanoribbon. As shown in Fig.~\ref{ff2}, we find that the edge-induced DMI vectors 
are slightly tilted from the direction orthogonal to the Cr-I-I-Cr plane, towards the out-of-plane direction of the monolayer. The intensity of the DMI vector between the Cr atoms at the edge is $0.87$ meV/atom, thus sizeable but several times smaller than in the case of iodine vacancy. Moreover, this DMI is truly localized at the edge, and vanishes immediately away from the edge, as pristine symmetry is restored. This is not surprising since the edge significantly affects the structural (and therefore the electronic) surrounding only of the Cr-Cr pair nearest to it, where Cr atoms lack the third Cr nearest-neighbor. It is interesting to note that in case of a zigzag edge the 2-fold rotational symmetry around horizontal direction is not removed, and this is reflected in the profile of DMI (see Fig.~\ref{ff2}). In the case of an armchair edge, we do not expect the structure remains stable since there will be four I atoms dangling at the outermost Cr-Cr pair. That would result in reconstruction or loss of I atom which is out of our scope in the present study.

\subsubsection{Chirality awaken by local strain} \label{strain} 

\begin{figure*}[ht]
\includegraphics[width=0.98\linewidth]{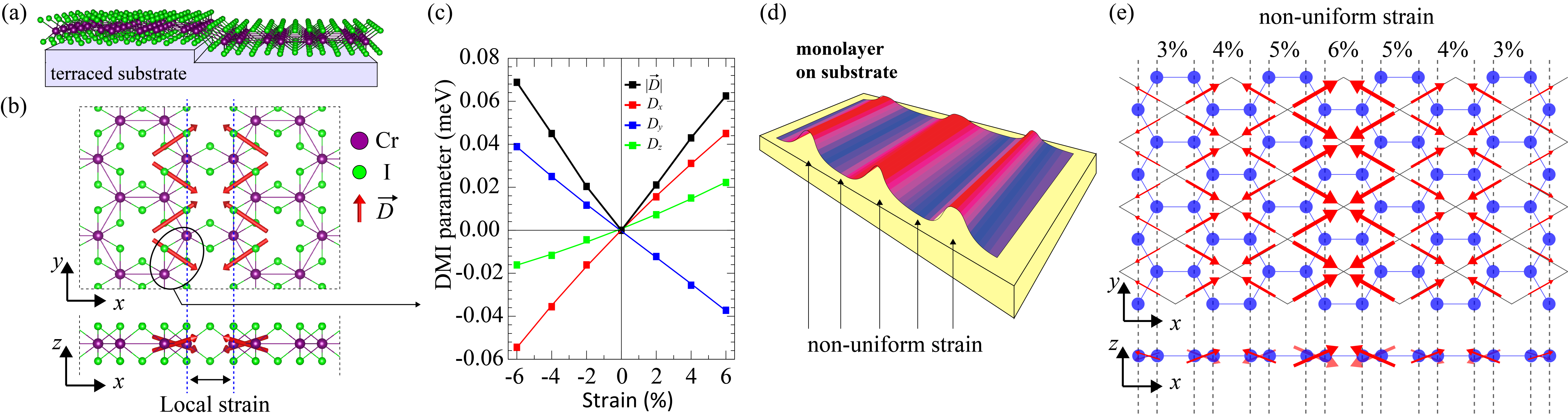}
\caption{\label{ff3}
\textbf{(a)} Schematic representation of a monolayer CrI$_3$ experiencing localized strain at the atomically-thin step between the terraces of the substrate. \textbf{(b)} Top view (above) and side view (below) of the profile of the awaken DMI (red arrows) in case of locally applied strain (mimicking the situation depicted in (a)). \textbf{(c)} Linear dependence between the components of awaken DMI and the localized strain. \textbf{(d)} The DMI pattern expected in the case of non-uniform distribution of strain in monolayer CrI$_3$, as found in e.g. ripples depicted on the side.}
\end{figure*}

Strain, either applied intentionally or induced locally due to interaction with a rough or terraced substrate (see Fig.~\ref{ff3}a), can strongly alter physical properties of a 2D crystal\cite{strain_images, strain_ref1, strain_ref2,PhysRevB.98.144411,deng2018strain}, including magnetic ones, and is often used to manipulate those properties in a controllable fashion. However, biaxial uniform strain does not break the inversion symmetry of a magnetic monolayer and will not cause appearance of DMI. Although less obvious, the same holds for uniaxial strain. Still, non-uniform straining is a well documented and a well established technique to manipulate e.g. optical properties of 2D materials, related to exciton confinement and single-photon emission, and can be employed to awaken DMI in magnetic 2D materials. The simplest scenario to consider the DMI due to symmetry broken by strain is the case of strain localized along an extended line, separating two parallel zig-zag chromium chains in a CrI$_3$ monolayer. As we schematically show in Fig.~\ref{ff3}a, such a case would be an idealized approximation of a monolayer laid over an atomic step between the flat terraces of the substrate, as often found at surfaces of e.g. SiC, MgO and other bulk crystals.

The blue dashed lines in Fig.~\ref{ff3}b indicate the strained section of the monolayer CrI$_3$ in our calculations. The released DMI vectors (shown by red arrows) are orthogonal to the bond between Cr atoms in the corresponding pair, pointing towards the strained section, and slightly tilted towards the in-plane of monolayer,  compared to direction orthogonal to the Cr-I-I-Cr plane. We examined the effects of both compressive and tensile strain, up to 6\% of the Cr-Cr distance in the pristine case. The obtained components of the DMI vector and its modulus are plotted as a function of applied strain in Fig.~\ref{ff3}c. The maximal realized DMI, corresponding to the 6\% of strain, amounted to 0.07 meV/atom. Nevertheless, it is quite remarkable (i) that DMI changes the direction with change of strain direction (stretch vs. compress), and (ii) that intensity of DMI and all its components change linearly with the applied strain, as shown in Fig.~\ref{ff3}c. These properties add to the versatility of realizable DMI patterns and the resulting spin textures in 2D magnetic materials, where linear dependence on strain is of great convenience for controlled manipulation. Bearing in mind the plethora of ways to generate nonuniform strain in 2D materials, our results suggest broad possibilities to control spin-chirality in a 2D system. For example, as depicted in Fig.~\ref{ff3}d, a gradient of strain experienced within a ripple of a 2D material would lead to a correspondingly varied DMI.

\subsubsection{Discussion} \label{disc}

Having considered a number of cases above to awake DMI, we attempt to quantify the emergent DMI by quantifying the break of the inversion symmetry using the electron density around the Cr-Cr pair. To this end, we define a quantity $\xi (\vec{R}_{0}, V_{D})\myeq{}\frac{1}{2}\int_{\vec{r}\in V_{D}}\!\lvert \rho(\vec{R}_{0} +\vec{r}) - \rho(\vec{R}_{0} -\vec{r})\rvert d^{3}\vec{r}$, where $\vec{R}_{0}$ is the position vector of the point of interest; $V_{D}$ stands for the domain of integration, which is chosen symmetrically around $\vec{R}_{0}$; $\rho$ is the valence electron charge density; and $\frac{1}{2}$ serves to avoid double-counting within the considered symmetric domain. We took $V_{D}$ as 0.2 \AA{} width shells around the point of interest, and enumerated such rings with integers $n = 1, 2, ..., n_{max}$ according to the proximity of the ring to the Cr-Cr pair. One expects that closer the symmetry breaking is to the Cr-Cr bond, it will yield larger influence on the emerging DMI. Hence, in order to quantify the backbone of DMI in this study,  we introduce quantity $\Omega (\vec{R}_{0},n_{max})=\sum_{n=1}^{n_{max}}\!\frac{1}{n}\xi (\vec{R}_{0}, n)$. By varying $n_{max}$ we have determined that the region that influences awakened DMI for a Cr-Cr pair of interest is bordered by the first passive ligand atoms, i.e. the nearest ligands that are not bridging respective Cr atoms. In Table \ref{omega}, we listed the obtained $\Omega$ values of the considered structures, next to the respectively found $|D|$. We note that $\Omega$ perfectly reflects the DMI in the system in case of structural deformations of CrI$_3$, where the iodine ligands of the Cr-pair are complete yet deformed for whatever reason. However, in case of vacancies or substitutions in the lattice, a need for an additional quantifier arises due to the strong changes in the spin-orbit coupling and the electronic structure in the vicinity of the Cr-pair.  

\begin{table}[ht]
    \centering
\begin{tabular}{|l|c|c|c|c|}
    \hline
    structure & pair & $\Omega (\vec{R}_{0},n_{max})$ & $|D|$ & $|D|\big/|J|$\\
              &      &              ($e$)             & (meV) &           \\
    \hline
    \hline
    Single vacancy & 1st pair & 1.44 & 4.68 & 0.39\\
                   & "inner" pair & 0.17 & 0.03 & 0.01\\
    \hline
    Line vacancies & 1st pair     & 1.69 & 5.57 & 0.48\\
                   & 2nd pair     & 1.18 & 2.47 & 1.06\\
                   & "inner" pair & 0.19 & 0.03 & 0.01\\
    \hline
    Edge           & edge pair    & 0.59 & 0.87 & 0.15\\
                   & "inner" pair & 0.18 & 0.00 & 0.00\\ 
    \hline
    Strain 6\%     & 1st pair & 0.31 & 0.08 & 0.02\\
    Strain 2\%     & 1st pair & 0.13 & 0.03 & 0.01\\
    \hline
    Sub. Br        & 1st pair & 0.74 & 0.09 & 0.025\\
    Sub. Cl        & 1st pair & 1.28 & 0.10 & 0.030\\
    Pseudosub Br-I        & 1st pair & 0.32 & 0.10 & 0.020\\
    \hline
\end{tabular}
    \caption{The calculated break of symmetry depending on the ground state electron charge density around the respective Cr-Cr pair, $\Omega (\vec{R}_{0},n_{max})$; modulus of the DMI vector, $|D|$; and the ratio of DMI and isotopic exchange interaction,  $|D|\big/|J|$.  }
    \label{omega}
\end{table}

As briefly mentioned in the section of Br and Cl substitution, the DMI interaction originates from the superexchange\cite{kanamori1959} between magnetic Cr cations, mediated by spin-orbit coupling on the bridging ligand. In the case of CrI$_{3}$, the hopping occurs between the $d$ orbitals of the nearest-neighbor $Cr^{3+}$ ions, through the conduction orbitals of bridging I atoms. Therefore, in order to clarify the DMI awaken in a system, one has to look at the electronic structure of the system, reflecting the different available hopping channels.

\begin{figure*}[ht]
\includegraphics[width=0.5\linewidth]{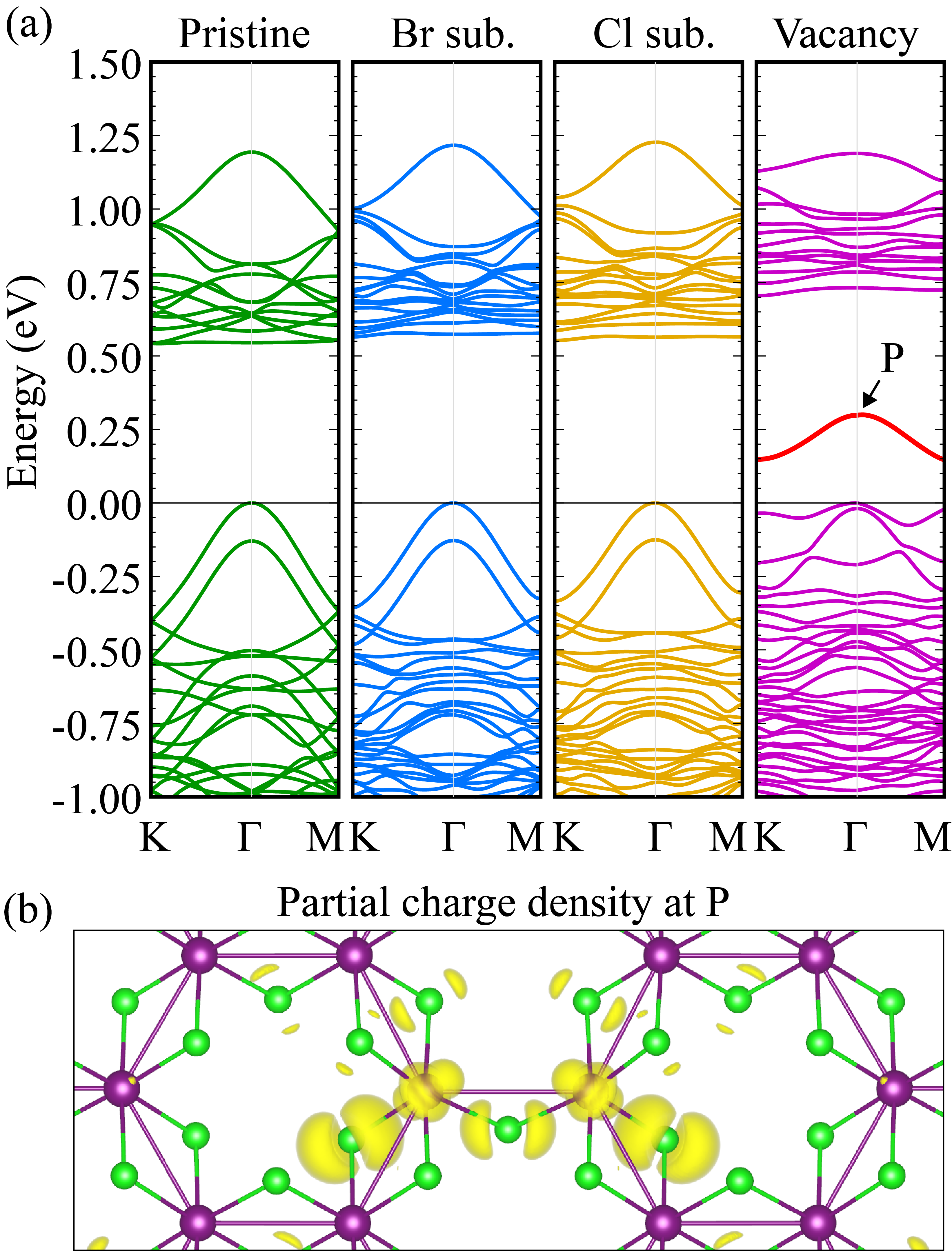}
\caption{\label{ff6}
\textbf{(a)} The electronic band structure of four different cases, pristine, single Br and Cl substitution cases, and vacancy defect. \textbf{(b)} The partial charge density of the midgap state at $\Gamma$ which we label by P. }
\end{figure*}

In Fig. \ref{ff6}a, we show the electronic band structure of the four different cases, pristine, Br and Cl substitutions and single I vacancy. First of all, it is clear that the single substitution of I by Br (blue) and Cl (orange) does not change electronic structure of the crystalline monolayer. One slight change is that the overall conduction band of Br and Cl is weakly shifted higher in energy. This can be associated with the overall decrease in the exchange interaction since slightly higher energies are needed for the excitations. In addition, there are small splittings of bands at the high symmetry points which appear as degenerate bands in the pristine case. These small variations of the electronic band structure in both cases of Br and Cl substitution compared to the pristine case make it reasonable to expect the small DMI as we obtained, even though our parametric break of spatial charge symmetry, $\omega$, is remarkably large in these cases.    

On the other hand, the band structure of the single iodine vacancy is completely different as compared to the pristine and substitution cases. Several occupied bands appear energetically closer to the Fermi level, and one midgap band is created just above the Fermi level. In order to understand the spatial character of the midgap band we calculated the partial charge density at the $\Gamma$ point, as shown in Fig.~\ref{ff6}b. Seemingly the absence of one I ligand creates an additional channel on the remaining I ligand that is energetically (strongly) favored for the exchange hopping from one Cr to another. Since that additional channel is fully asymmetric with respect to the Cr-Cr pair, it is the source of the large DMI seen in the case of an iodine vacancy.

\subsubsection{Chirality versus the electric field}

\begin{figure*}[ht]
\includegraphics[width=0.98\linewidth]{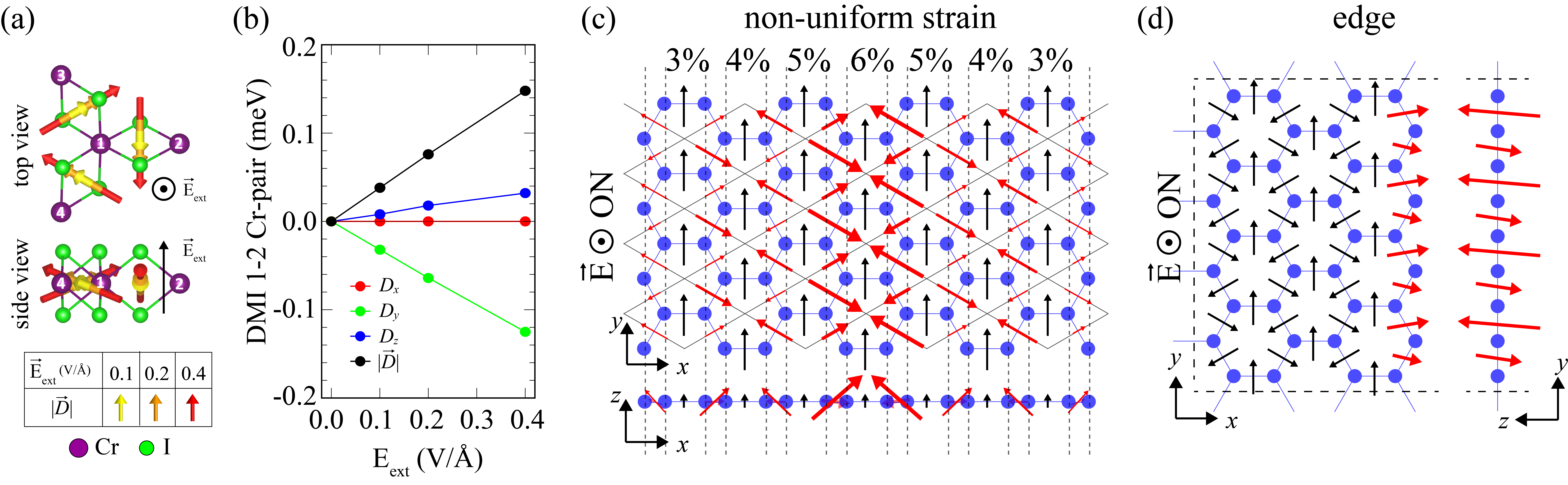}
\caption{\label{ff4}
\textbf{(a)} The profile of awaken DMI in case of externally applied electric field on the monolayer CrI$_{3}$, top (above) and side view (below). \textbf{(b)} Linear dependence between the components of awaken DMI and applied electric field. \textbf{(c)} The schematic representation of the DMI in case of non-uniformly applied strain, together with the externally applied electric field orthogonal to the monolayer. \textbf{(d)} The schematic representation of the DMI in case of the edge influence and external electric field.}
\end{figure*}

It was recently shown that external out-of-plane electric field breaks the inversion symmetry in monolayer ferromagnets, and induces DMI - growing linearly with increasing applied electric field \cite{EfieldDMI,EfieldDMI1,EfieldDMI2,EfieldDMI3}. However, the intensity of DMI in monolayer CrI$_{3}$ is still under debate, since the reported values strongly vary from one article to another, for example from 0.01 to 0.4 meV/atom for electric field of 0.1 V/\AA. It is likely that such large variation in reported DMI values stems from the choices of exchange-correlation functional approximations in respective DFT calculations \cite{EfieldDMI3}. 

In order to compare the results to DMI released at the structural defects presented so far, we performed the calculations of DMI in a CrI$_{3}$ monolayer exposed to electric field using the consistent methodology throughout all the calculations. In Fig.~\ref{ff4}a we show top and side view of the DMI vectors induced in the CrI$_{3}$ monolayer for the applied out-of-plane electric field of 0.1, 0.2 and 0.4 V/\AA, beyond which value the structure was no longer stable. Note that although the inversion symmetry is broken between the nearest-neighbor Cr atoms when electric field is introduced, the mirror plane orthogonal to their bond still defines the symmetry operations. This results in the DMI vector orthogonal to the bond, as restricted by the Moriya's symmetry rules \cite{dmi_moriya}. This can be seen in Fig.~\ref{ff4}b, where $x$-component of the DMI vector (parallel to the corresponding Cr-Cr bond) remains consistently zero, as the consequence of symmetry rules. On the other hand, the $y$ and $z$ DMI components, as well as the total DMI intensity, grow linearly with the external electric field. The increase of electric field by 0.1 V/\AA results in 0.04 meV gain in total DMI per Cr-pair. Moreover, DMI induced by gating will change the sign with the change of polarity of the applied electric field. Therefore, as was case with the locally applied strain, external electric field can be used as the turning knob to modify both magnitude and direction of the spin-chirality in a magnetic 2D material. 

Since local strain and external electric field release comparable DMI, it is instructive to combine the inhomogeneous strain with the out-of-plane external electric field to achieve richer patterns of possible chiral spin textures. In Fig.~\ref{ff4}c we present a DMI profile that arises from nonuniform strain as shown in Fig.~\ref{ff3}d when exposed to external electric field. One sees that that DMI vectors induced in two manners `interfere' in constructive and destructive way intermittently on the lattice, where DMI will be amplified by external electric field on one checkerboard sublattice of Cr-Cr bonds and reduced on the other. In addition, with inclusion of the external electric field significant DMI will arise even on Cr-pairs where strain did not induce any (depicted by black arrows in Fig.~\ref{ff4}c). 

DMI pattern due to a joint effect of the sample edge and the applied electric field is depicted in Fig.~\ref{ff4}d. Similarly to the previous case, the electric field will intermittently amplify and reduce the DMI vectors present at the edge, and release additional DMI further from the edge. However, here the influence of external field will be smaller compared to the case of inhomogeneous strain, since DMI induced by the symmetry breaking on edge is significantly larger than the maximal one induced by electric field.

\begin{figure*}[ht]
\includegraphics[width=0.98\linewidth]{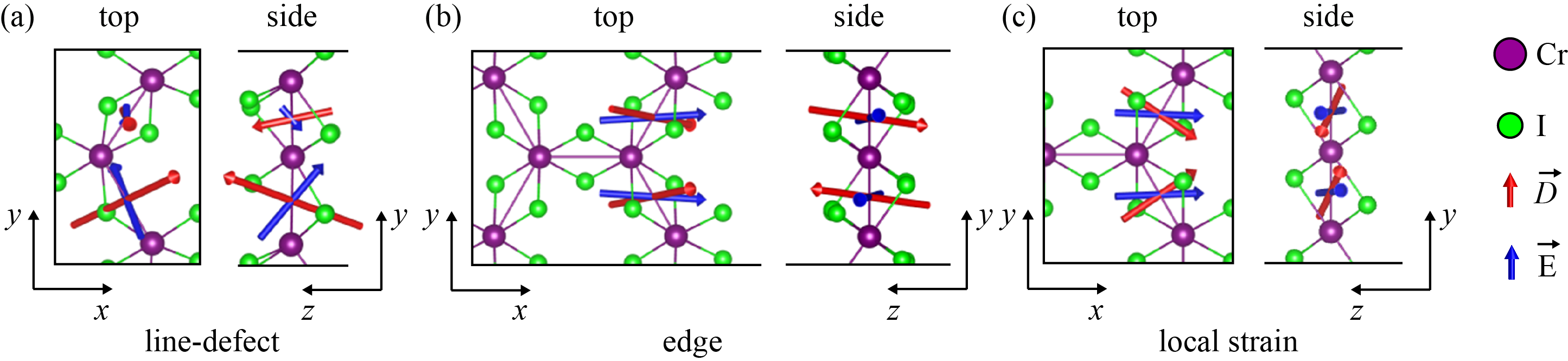}
\caption{\label{ff5}
\textbf{(a)} The released DMI versus the intrinsic electric field generated by the symmetry breaking in monolayer CrI$_3$, for all investigated structures. \textbf{(b)} The geometric relation between the vectors of DMI and of the intrinsic electric field, for three studied structural deviations from the pristine crystal.}
\end{figure*}

In the previous sections we have shown how different ways of inversion symmetry breaking induce different profiles of DMI. Different scales for DMI values suggest that in case of physically removed atoms from monolayer, inversion symmetry is 'more' broken than in case of inhomogeneous periodic strain. Therefore, it would be very useful to quantify this degree of inversion symmetry breaking between nearest neighbor Cr atoms. 
In case of pristine crystal, next to DMI in the point between Cr atoms, total crystal electric field is 0 as well, due to inversion symmetry. Moreover, net-average of this electric field around that point is 0 as well. Next to this, introduction of electric field breaks the symmetry, as discussed. Therefore, it is reasonable to assume the vice versa: that when inversion symmetry is broken between Cr-Cr nearest neighbor pair, besides DMI, significant net crystal electric field will appear as well. We will use this field to quantify the break of inversion symmetry. In order to calculate the crystal electric field in the point between two Cr atoms, for different structures, we take into account the charge density of (valence) electrons as the $\omega$ calculation.

In Fig.~\ref{ff5}a we present the intensity of the crystal electric field between the Cr-Cr pairs of interest, for each structure under investigation, together with the corresponding DMI. In Fig.~\ref{ff5}b one can see the relative position of the two vectors. The results presented here not only confirm that DMI arises with the break of inversion symmetry (non-zero electric field), but also provide an additional information, about the degree of symmetry breaking with different deviations from pristine monolayer. Applied strain, vacancy defects and edge of a monolayer can be understood as a building block for awakening DMI by breaking the symmetry, each block having its particular characteristics. Locally applied strain breaks the inversion symmetry on a smaller scale, just as external field. This is reflected in the electric field between two nearest neighbor Cr atoms of order of magnitude 0.1 V/\AA $ $ and the consequence is DMI intensity of 0.01-0.1 meV/atom, tunable with the strain. Note that we report the same DMI and electric field between Cr-Cr pair in case of 6\% local strain and 0.2 V/\AA external electric field, applied to the monolayer CrI$_{3}$. Contrary to the local strain, or external electric field, edge or vacancy defects break the symmetry on the bigger scale. In this case, the good scale for the electric field is 1 V/\AA, and awaken DMI has value of 1-6 meV/atom. In order to fully characterize each 'building block' for releasing DMI, we present the DMI and inner electric field vectors, between Cr-Cr pairs on Fig.~\ref{ff5} (d) for each structural deviation from pristine crystal. 

One aspect of the results presented here lies in the fact that size of electric field induced by different structural deviations, corresponds to the size of awaken DMI. This suggests that calculated electric field in-between two Cr atoms is a good measure for the degree of symmetry breaking and therefore, it is reasonable to perform atomic-scale electric field measurements with each successful formation of magnetic monolayer. These measurements will contain information about the position and degree of the symmetry break in a monolayer, and therefore implicitly, information about the position and magnitude of released DMI in that monolayer.

Another aspect of our results comes from the fact that released DMI is restricted to only few \AA $ $ from the structural deviation that causes it, for each building block. This further suggests that effect of two or more blocks combined can be considered as a superposition of individual effects, without significant `interaction' of the blocks. Therefore, one can now estimate and even design many different profiles of DMI, as we did in a few examples, by combining the building blocks presented here.\footnote{Though we restrict our analysis on typical example of monolayer ferromagnet, CrI$_{3}$, there is no reason why all conclusions from here could not be applied on monolayer antiferromagnets, such as e.g. FePS$_{3}$.} 

In summary, we have brought out first guidelines on how to release DMI in 2D magnetic materials and tailor its textures further. Related to the amount of released DMI, we also quantified the inversion symmetry breaking in monolayer CrI$_{3}$, after calculating the emergent electric field inside the crystal in cases of different structural deviations from the pristine crystal. Moreover, we have compared the obtained results with the DMI awaken by external electric field, and discussed the cumulative effects of different DMI sources. The employed principles in this work remain valid in any emergent magnetic 2D material, and our results provide clear yet rich pathways towards design, control and characterisation of versatile DMI and spin textures and their use in spintronic and magnonic applications of magnetic monolayers.


\begin{acknowledgement}

This work was supported by the Research Foundation-Flanders (FWO-Vlaanderen) and the Special Research Funds of the University of Antwerp (TOPBOF). The computational resources and services used in this work were provided by the VSC (Flemish Supercomputer Center), funded by Research Foundation-Flanders (FWO) and the Flemish Government -- department EWI.

\end{acknowledgement}

\bibliography{achemso-demo}

\end{document}